\documentclass[a4paper]{article}

\usepackage{INTERSPEECH2020}
\usepackage{multirow}

\title{Leveraging Text Data Using Hybrid Transformer-LSTM Based End-to-End ASR in Transfer Learning}
\name{%
\parbox{0.9\linewidth}{\centering
      Zhiping Zeng$^{1,2}$,
      Van Tung Pham$^{1}$,
      Haihua Xu$^1$,
      Yerbolat Khassanov$^{1,3}$,
      Eng Siong Chng$^1$,
      Chongjia Ni$^4$ and
      Bin Ma$^4$%
    }%
}
\address{
$^1$School of Computer Science and Engineering, Nanyang Technological University, Singapore\\
$^2$Huya AI, Guangzhou, China\\  
$^3$ISSAI, Nazarbayev University, Kazakhstan\\
$^4$Machine Intelligence Technology, Alibaba Group}

\begin{document}

\maketitle

\begin{abstract}
In this work, we study leveraging extra text data to improve low-resource end-to-end ASR under cross-lingual transfer learning setting. 
To this end, we extend our prior work \cite{Tung2020}, and propose a hybrid Transformer-LSTM based architecture.
This architecture not only takes advantage of the highly effective encoding capacity of the Transformer network but also benefits from extra text data due to the LSTM-based independent language model network.
We conduct experiments on our in-house Malay corpus which contains limited labeled data and a large amount of extra text. 
Results show that the proposed architecture outperforms the previous LSTM-based architecture \cite{Tung2020} by 24.2\% relative word error rate (WER) when both are trained using limited labeled data. 
Starting from this, we obtain further 25.4\% relative WER reduction by transfer learning from another resource-rich language. 
Moreover, we obtain additional 13.6\% relative WER reduction by boosting the LSTM decoder of the transferred model with the extra text data. 
Overall, our best model outperforms the vanilla Transformer ASR by 11.9\% relative WER. 
Last but not least, the proposed hybrid architecture offers much faster inference compared to both LSTM and Transformer architectures.

\end{abstract}

\section{Introduction}
End-to-end (E2E) architecture has been a promising strategy for ASR. In this strategy, a single network is employed to directly map acoustic features into a sequence of characters or subwords without the need of a pronunciation dictionary that is required by the conventional hidden Markov model based systems. Furthermore, the components of the E2E network can be jointly trained using a common objective criterion to achieve overall optimization which greatly simplifies the ASR development process. Although the simplicity of E2E ASR architecture is attractive, especially for new languages, it requires a huge amount of labeled training data. 

In this work, we focus on E2E ASR for a low-resource language.
Specifically, we assume that the target language possesses a limited amount of labeled data to train E2E systems, while an extra text corpus of the language can be easily collected.
Additionally, we assume that we possess a large amount of labeled data from another resource-rich source language.
This is a common scenario in real-world applications.

The extra text is usually employed to train language models (LM) applied during decoding \cite{Graves:2014:TES:3044805.3045089,Hori2018,deepAndShallowFusion} and re-scoring stages \cite{ChanJLV16}. Such techniques not only require external language models but also
lead to a slow inference. To tackle this problem, \cite{Tung2020} has proposed long short term memory (LSTM)-based encoder-decoder architecture which allows improving the LM capacity of the decoder using the extra text data. However, it utilized the LSTM structure for the encoder which has shown limited modeling capacity as well as slow training. On the other hand, Transformer \cite{Transformer2017} has been a promising approach for E2E ASR due to its high modeling capacity and fast training. However, its decoder closely interacts with the encoder output through an encoder-decoder cross-attention. Therefore, it is not straightforward to employ extra text data to improve the decoder. 

In this work, we propose a hybrid Transformer-LSTM architecture which combines the advantages of \cite{Tung2020} and \cite{Transformer2017}. It not only has a high encoding capacity of the Transformer but also benefits from the extra text data due to the LSTM-based independent language model decoder.
To further benefit from the labeled data from another language, we employ cross-lingual transfer learning, which is a popular approach to address the limited resource problem in ASR \cite{Huang2013,Heigold2013,Shan2019,Tong2017,Dalmia2018}, on the proposed architecture.
Specifically, we first use labeled data of the resource-rich language to train an ASR model and then transfer it to the target language.
Lastly, the extra text data is used to boost the decoder of the transferred model. 



The paper is organized as follows. Section \ref{trainE2E} describes baseline architectures mentioned in \cite{Tung2020} and \cite{Transformer2017}. Then, the proposed techniques are presented in Section \ref{proposedArchitecture}. Experimental setup and results are presented in Section \ref{expSetup} and \ref{expResults} respectively. Section \ref{conclusions} concludes our work.

\vspace{-0.2cm}
\section{Baseline architectures}
\label{trainE2E}
\subsection{LSTM-based encoder-decoder architecture}
\label{ILM}
\begin{figure}[h]
  \centering
\includegraphics[scale=0.35]{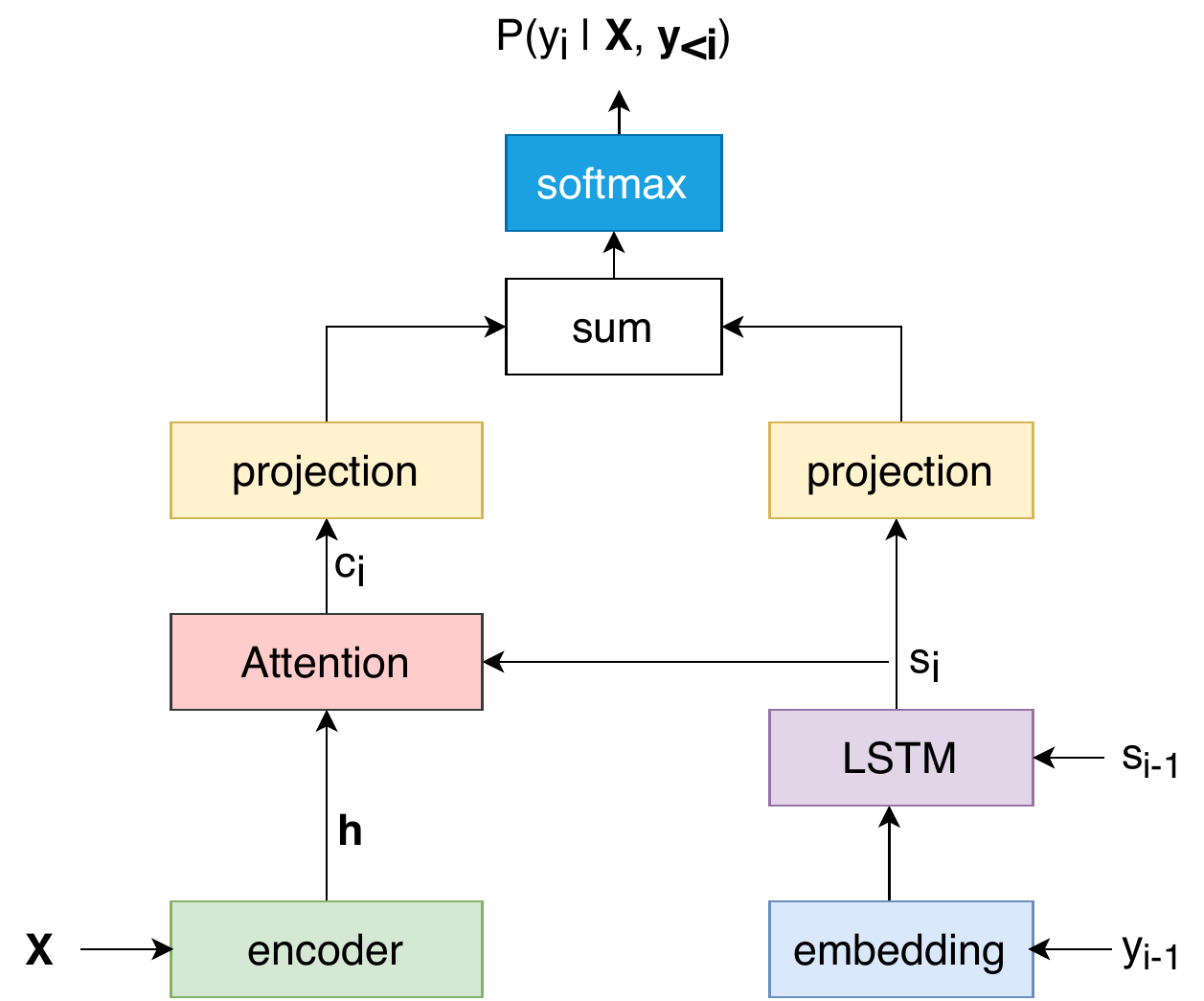}
  \caption{LSTM-based encoder-decoder architecture ($A_1$) \cite{Tung2020}, where the decoder acts as an independent language model.}
  \label{s2cModel}
\end{figure}

A LSTM-based encoder-decoder architecture \cite{Tung2020}, denoted as $A_1$ in the rest of this paper, consists of a Bidirectional LSTM encoder and a LSTM-based decoder which are shown in Fig. \ref{s2cModel}. Let $<$\textbf{X}, \textbf{Y}$>$ be a training utterance, where $\textbf{X}$ is a sequence of acoustic features and $\textbf{Y} = \{y_1, y_2,..., y_{|\textbf{Y}|} \}$ is a sequence of output units. The encoder acts as an acoustic model which maps acoustic features $\textbf{X}$ into an intermediate representation \textbf{h}. Then, the decoder, which consists of an embedding, a LSTM and a projection layers, generates one output unit at each decoding step $i$ as follows,
\begin{align}
    s_i &= LSTM(s_{i-1}, embedding(y_{i-1})) \label{decoderNew}\\
    c_i &= attention(\textbf{h}, s_{i}) \label{attentionNew}\\
        P(y_i \mid \textbf{X}, y_{<i}) =\ & softmax(proj(s_i) +  proj(c_i)) \label{distrNew}
\end{align}
where $c_i$ is the context vector, $s_{i-1}$ and $s_{i}$ are output hidden states at time step $i-1$ and $i$ respectively, $embedding()$ and $proj()$ are embedding and projection layers respectively.

From Equation \eqref{decoderNew}, the LSTM is only conditioned on the previous decoding hidden state and previous decoding output. In other words, the LSTM acts as an independent language model that can be easily updated with text-only data \cite{Tung2020}. 

\subsection{Transformer encoder-decoder architecture}
\label{transf}
\begin{figure}[h]
  \centering
\includegraphics[scale=0.35]{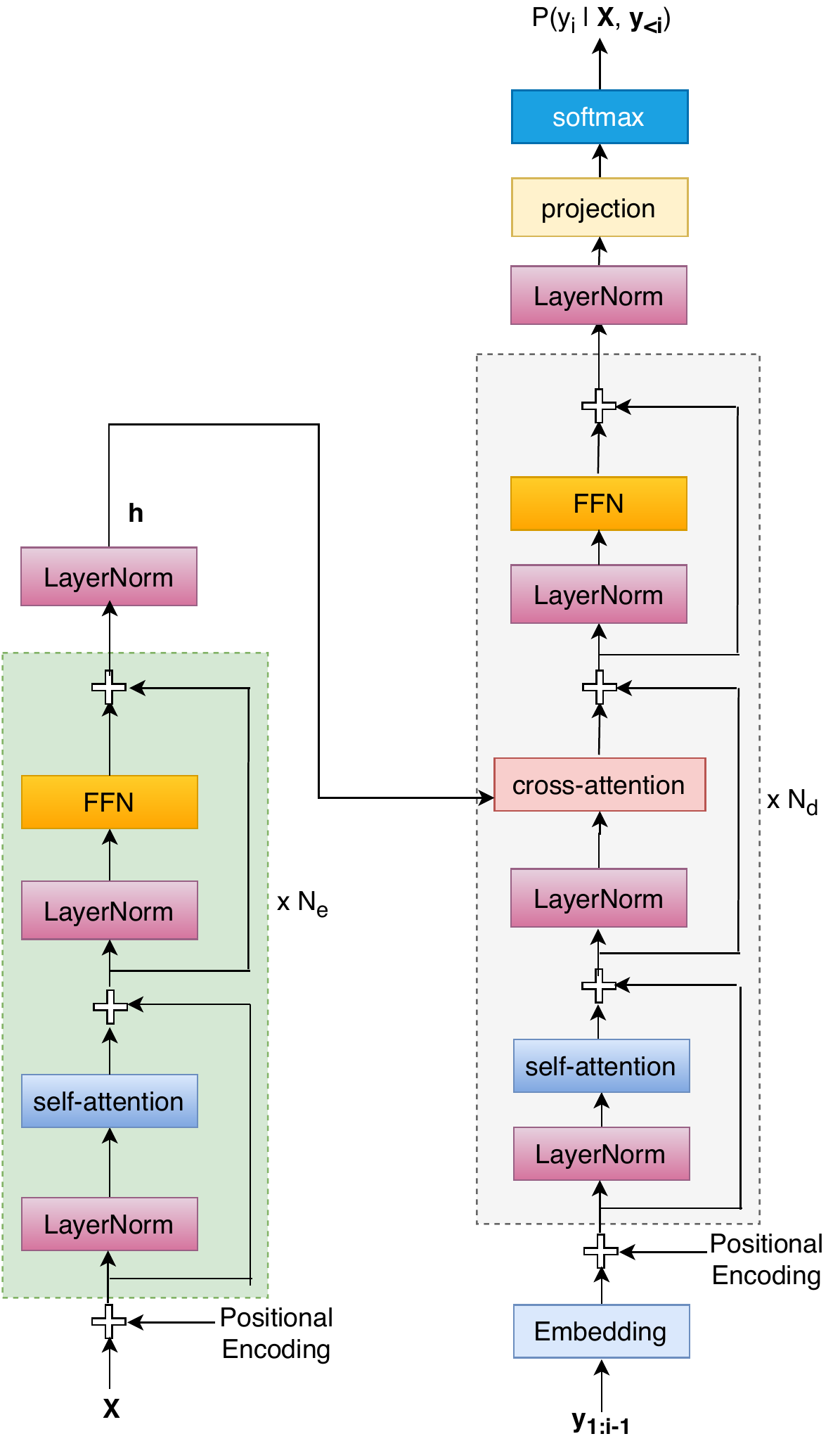}
  \caption{Transformer architecture ($A_2$).}
  \label{TransFig}
\end{figure}

Transformer has been proposed in \cite{Transformer2017} for sequence-to-sequence modeling in natural language processing tasks, then adopted to the ASR task in \cite{speechTrans,TransASRU}. The model architecture, denoted as $A_2$, is shown in Fig. \ref{TransFig}.  

The encoder is shown in the left half of Fig. \ref{TransFig}. It consists of $N_e$ encoder-blocks, each of them has two sub-blocks: self-attention and position-wise feed-forward network (FFN). The self-attention employs multi-head attention (MHA) which is a function of three inputs: query $Q$, key $K$ and value $V$. It includes multiple heads which can be processed in parallel. Each head employs a dot-product attention ($DotProdAtt$) as follows,  
\begin{equation}\label{dotProd}
    DotProdAtt(Q,K,V) = softmax(\frac{QK^T}{d_k} )V
\end{equation}
where $d_k$ is the hidden dimension. Besides, layer normalization \cite{BaKH16} and residual connection \cite{HeZRS16} are introduced to each encoder-block for effective training.

The decoder is shown in the right half of Fig. \ref{TransFig} which consists of $N_d$ decoder-blocks. Different from the encoder-block, each decoder-block has one more sub-block, i.e. the cross-attention. Its first input comes from the previous sub-block output while another input is from the output of the encoder.   

\section{Proposed techniques}
\label{proposedArchitecture}
To exploit extra text data while having high modeling capacity, we propose a Transformer-LSTM architecture in Section \ref{TransEnc}.
Then, we exploit using extra text data to improve the proposed architecture under the cross-lingual transfer learning setting.

\subsection{The Transformer-LSTM architecture}
\label{TransEnc}
In this section, we first compare approaches presented in Section \ref{ILM} and \ref{transf}. Then, based on the comparison, we propose a novel architecture that takes advantage of both approaches. 

Previous work \cite{combineTransformerLSTM} showed that the Transformer not only produces better encoding representation but is also faster than the LSTM counterpart in training. First, the Transformer encoder uses dot-product attention (see Equation (\ref{dotProd})) which allows each position has access to information from all other positions in the same sequence regardless of their distance. In contrast, although in theory LMST can model long-range dependence, in practice it faces difficulty to capture dependencies of far-distance elements \cite{TangMRS18} which limits its modeling capacity for long sequences such as acoustic signal.
Second, by relying entirely on feed-forward components, the Transformer model avoids any sequential dependencies, and hence can maximize parallel training. In contrast, training a LSTM-based network is slow due to the recurrence property of the LSTM. 

Despite being highly effective, the Transformer decoder is not easy to be improved using text-only data. Specifically, the decoder includes the cross-attention sub-block which is conditioned on the encoder output. In contrast, the LSTM-based decoder (in Section \ref{ILM}) can be easily boosted using the text data.
Another issue of the Transformer decoder is slow inference \cite{zhang-Etal:2018:ACL2018accelerating}. Specifically, to generate an output $y_i$, the decoder needs to process all previous decoding units $y_{1:i-1}$.  On the other hand, the LSTM-based decoder has faster inference since it only needs the last output unit $y_{i-1}$ to generate $y_i$.

Based on the above comparisons, we propose a hybrid architecture that takes advantage of both Transformer and LSTM
architectures. Specifically, our encoder is from Transformer, while the decoder is taken from the LSTM architecture. The benefits of the proposed architecture lie in two aspects. First, it has high modeling capacity, as well as faster training and decoding. Second, the LSTM-based architecture allows us to easily leverage text data to boost the decoder, yielding improved ASR performance. We denote the proposed architecture as $A_3$ in the rest of this paper.

\subsection{Exploiting extra text data under cross-lingual transfer learning}
\label{useExtraText}
\begin{figure}[h]
  \centering
  \includegraphics[scale=0.4]{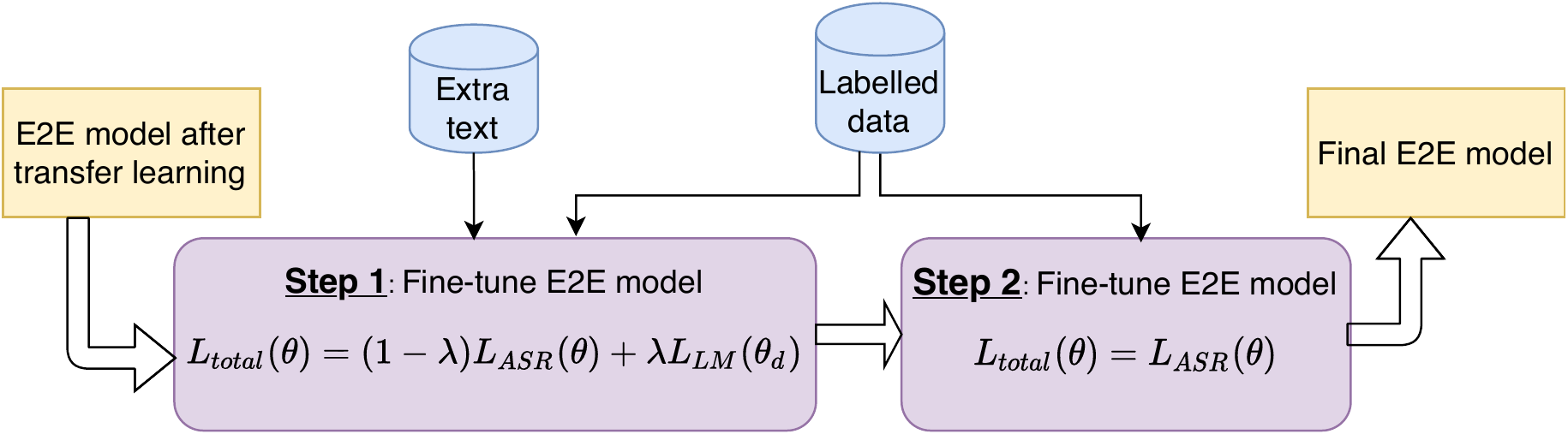}
  \caption{Boosting the transferred model using extra text data of the target language.}
  \label{fine_tune}
\end{figure}

To tackle the low-resource training problem, we first perform cross-lingual transfer learning. We start with training E2E models using a source language. We then replace the language-dependent components of the decoder (i.e. the embedding and output projection layers) of the source language by those of the target language. Finally, 
the models are fine-tuned using labeled data of the target language. Although transfer learning is not our focus, to achieve the best performance, we carefully examine various transfer settings as presented in Section \ref{exp:resTransfer}. More importantly, we aim to boost the transferred model of the proposed architecture using extra text data of the target language. Fig. \ref{fine_tune} describes our process. 

From Fig. \ref{fine_tune}, the entire process is implemented with two main steps. In the first step, we merge the extra text and the labeled data together to fine-tune the transferred model. This avoids a so-called catastrophic forgetting problem as mentioned in \cite{Tung2020}. Specifically, at each training iteration, we mix a batch of labeled data consisting of $B_{labeled}$ utterances with a batch of text data consisting of $B_{text}$ utterances to fine-tune the transferred model with the following loss function:
\begin{equation}\label{totalLoss}
    L_{total}(\theta) = (1 - \lambda) L_{ASR}(\theta)  +  \lambda L_{LM}(\theta_{d})
\end{equation}
where $\lambda$ denotes an interpolation factor, $\theta$ and $\theta_d$ denote entire E2E parameters and decoder parameters respectively, $L_{ASR}(\theta)$ and $L_{LM}(\theta_{d})$ denote the ASR loss and LM loss generated by the labeled data and text data respectively.
In the second step, the model is further fine-tuned with the labeled data of the target language. Similar to \cite{Tung2020}, we empirically found that the second step is necessary to improve overall performance. 
\section{Experimental setup}
\label{expSetup}

\subsection{Data}
\label{exp:Data}
We conduct experiments on our in-house corpus of Malay language which consists of limited labeled data plus extra text. We split the labeled data into three sets for training, validation and evaluation.
Detailed division is shown in Table \ref{exp:corpus}. We perform speed perturbation based data augmentation \cite{DBLP:conf/interspeech/KoPPK15} on the training data. For extra text data, we examine two sets: the first set has over 8 million sentences, denoted as $T1$; while the second set is a subset of the first one which consists of 2 million sentences, denoted as $T2$.

For the source language, which is English, we use two subsets of the National Speech Corpus (NSC)~\cite{koh2019building} to train source models. The first subset, denoted as $S_{200h}$ consists of 200 hours, where we also apply speed perturbation based data augmentation. The second subset consists of 1000 hours, denoted as $S_{1000h}$. 
\vspace{-0.2cm}
\begin{table}[htbp]
\small
\caption{Detailed division of the labeled Malay data.}
\centering
\begin{tabular}{l|c|c|c}
\hline
\multicolumn{1}{l|}{} & \multicolumn{1}{l|}{Train} & \multicolumn{1}{l|}{Dev} & \multicolumn{1}{l}{Test} \\ \hline
\#Speakers & 57    & 6   & 6    \\
\#Utterances             & 8500                       & 1785                     & 1957                      \\
Length (hours)    & 20.6  & 4.3 & 4.1 \\ \hline
\end{tabular}
\label{exp:corpus}
\end{table}

\subsection{E2E setting}
ESPnet toolkit \cite{watanabe2018espnet} is used to train our E2E architectures. We use 80 mel-scale filterbank coefficients with pitch as input features, and 500 Byte-Pair Encoding (BPE) units are used as output units. For all E2E architectures, the acoustic features are processed by the VGG network \cite{kim2017joint}. Detailed setting of architectures can be seen in Table  \ref{exp:setting}. Each BLSTM layer has 320 cells, while each LSTM layer of the decoder has 256 cells. In each self-attention and cross-attention sub-blocks, we use 4 heads with hidden dimension of 2048. The FFN consists of two ReLu activation functions and two affine transforms with size of 2048. To allow transfer learning, we use the same settings for both source and target languages. During the fine-tuning process in Section \ref{useExtraText}, we set $B_{labeled} = 30$ and $B_{text} = 90$.
\begin{table}[htbp]
\small
\setlength{\tabcolsep}{4pt}
\centering
\caption{Setting of different E2E architectures.}
\begin{tabular}{c|l|l|l}
\hline
\multicolumn{1}{l|}{} & \multicolumn{1}{l|}{ \# Paras} & \multicolumn{1}{l|}{encoders} & \multicolumn{1}{l}{decoders} \\ \hline
$A_1$                     & 77.4M                                                                              & 6 BLSTM                         & 1 LSTM                         \\
$A_2$                     &  120 M                                                                     & 12 (self-att + FFN)                        &  6 (self-att + \\                     & & & cross-att + FFN)\\
$A_3$                     &  81.1M                                                                     & 12 (self-att + FFN)                        & 1 LSTM                             \\
\hline              
\end{tabular}
\label{exp:setting}
\end{table}
\vspace{-0.2cm}
\section{Results and analysis}
\label{expResults}
The overall ASR performance on the test set after applying proposed techniques (in Section \ref{proposedArchitecture}) is presented in Table \ref{exp:results}. In following subsections, we will describe and elaborate results for each proposed technique.
\begin{table}[htbp]
\centering
\caption{The WER(\%) on the test set of different E2E models. $S_{1000h}$-Encoder denotes that we use the data $S_{1000h}$ of the source language to train a source model, then transfer only the encoder of the source model to the target language. $T1$ denotes an extra text corpus of the target language that consists of 8 million sentences.}
\setlength{\tabcolsep}{1.1mm}
\begin{tabular}{c|c|c|c|c}
\hline
No. & Architecture & {\begin{tabular}[c]{@{}c@{}}Transfer learning\\ settings \end{tabular}}  & Extra text     & {WER(\%)}      \\ \hline
1                 & $A_1$             & -            & -               & 18.2 \\ \hline
2                 & $A_2$      & -            & -               & 12.3 \\ \hline
3                 & $A_3$ & -            & -               & 13.8 \\ \hline \hline
4            & $A_2$      & \multicolumn{1}{l|}{$S_{1000h}$-Encoder} & -               & 10.1 \\ \hline
5             & $A_3$     & \multicolumn{1}{l|}{$S_{1000h}$-Encoder} & -               & 10.3 \\ \hline 
6             & $A_3$     & \multicolumn{1}{l|}{$S_{1000h}$-Encoder} & +\ $T1$               & \textbf{8.9} \\ \hline
\end{tabular}
\label{exp:results}
\end{table}
\vspace{-0.2cm}
\subsection{Results of different E2E architectures}
This section presents the results of different E2E architectures, i.e. $A_1$, $A_2$ and $A_3$ trained using only the labeled data of the target language. The word error rate (WER) results are reported in the first three rows in Table \ref{exp:results}. We also report the decoding speed of these models in Table \ref{exp:compareArchitectures}. As we can see, $A_3$ not only outperforms $A_1$ by 24.2\% relative WER but also offers 1.5 times faster decoding. This indicates that employing the Transformer for the encoder is very effective. $A_2$ achieves best WER, but has much slower decoding speed compared to $A_3$. 
\vspace{-0.2cm}
\begin{table}[htbp]
\setlength{\tabcolsep}{4pt}
\centering
\caption{Decoding speed of different E2E architectures.}
\begin{tabular}{c|c}
\hline
\multicolumn{1}{l|}{Architecture} &  \multicolumn{1}{l}{\begin{tabular}[c]{@{}c@{}}Decoding speed\\ (seconds/utt) \end{tabular}}\\ \hline
$A_1$   & 7.5   \\ 
$A_2$   & 31.61   \\  
$A_3$   & 5.00   \\ \hline 
\end{tabular}
\label{exp:compareArchitectures}
\end{table}
\vspace{-0.2cm}
\subsection{Results of transfer learning}
\label{exp:resTransfer}

In this section, we examine the ASR performance of different transfer learning settings.
Firstly, we examine the effect of using different amounts of source language data on the target language's performance. Secondly, we analyze the different level of transfer learning: (1) only $l$ (e.g. $l$ = 3, 6, 9) bottom layers of the encoder is transferred; (2) entire encoder is transferred; (3) both encoder and decoder (except embedding and projection layers) are transferred. 
These experiments were conducted only for $A_2$ and $A_3$, since $A_1$ produced worst results.
The experiment results are given in Table \ref{exp:layers}. 

We observed that transferring the entire encoder achieves the best results for both $A_2$ and $A_3$. Additionally, transfer with 1000 hours of source data is noticeably better than that of 200 hours with data augmentation. The best results are summarized in rows 4 and 5 in Table \ref{exp:results}. It can be seen that cross-lingual transfer learning leads to significant improvement for both $A_2$ and $A_3$. For example, the result in row 5 outperforms that of row 3 by 25.4\% relative WER (from 13.8\% to 10.3\%).

\begin{table}[htbp]
\small
\centering
\caption{ASR performance of different transfer learning settings.}
\begin{tabular}{c|c|c|c|c}
\hline
\multicolumn{1}{l|}{\multirow{2}{*}{}} & \multicolumn{1}{l|}{\multirow{2}{*}{Source data}} & \multicolumn{1}{c|}{\multirow{2}{*}{Transfer modules}} & \multicolumn{2}{c}{WER(\%)}                             \\ \cline{4-5} 
\multicolumn{1}{l|}{}                  & \multicolumn{1}{l|}{}                        & \multicolumn{1}{l|}{}                                    & \multicolumn{1}{c|}{Dev} & \multicolumn{1}{c}{Test} \\ \hline
\multirow{5}{*}{$A_2$}      & \multirow{5}{*}{$S_{200h}$}      & Encoder + Decoder  & 15.1 & 11.0 \\ \cline{3-5} 
                         &                               & Encoder      & 14.9 & 10.4 \\ \cline{3-5} 
                         &                               & 9 bottom encoder layers     & 16.2 & 11.6 \\ \cline{3-5} 
                         &                               & 6 bottom encoder layers     & 16.9 & 12.3 \\ \cline{3-5} 
                         &                               & 3 bottom encoder layers     & 17.3 & 12.6 \\ \hline
\multirow{5}{*}{$A_2$} & \multirow{5}{*}{\textbf{$S_{1000h}$}} & Encoder + Decoder         & 14.2 & 10.4 \\ \cline{3-5} 
                         &                               & \textbf{Encoder}      & \textbf{13.9} & \textbf{10.1} \\ \cline{3-5} 
                         &                               & 9 bottom encoder layers     & 15.3 & 11.1 \\ \cline{3-5} 
                         &                               & 6 bottom encoder layers     & 16.7 & 12.3 \\ \cline{3-5} 
                         &                               & 3 bottom encoder layers     & 17.1 & 12.7 \\ \hline \hline
                         
\multirow{5}{*}{$A_3$}      & \multirow{5}{*}{$S_{200h}$}      & Encoder + Decoder  & 15.5 & 11.0 \\ \cline{3-5} 
                         &                               & Encoder         & 15.1 & 10.8 \\ \cline{3-5} 
                         &                               & 9 bottom encoder layers     & 16.4 & 12.2 \\ \cline{3-5} 
                         &                               & 6 bottom encoder layers     & 17.4 & 12.8 \\ \cline{3-5} 
                         &                               & 3 bottom encoder layers     & 17.6 & 13.1 \\ \hline
\multirow{5}{*}{$A_3$} & \multirow{5}{*}{$S_{1000h}$} & Encoder + Decoder       & 14.6  & 10.7 \\ \cline{3-5} 
                         &                               & \textbf{Encoder}      & \textbf{14.3} & \textbf{10.3} \\ \cline{3-5} 
                         &                               & 9 bottom encoder layers     & 15.5 & 11.3 \\ \cline{3-5} 
                         &                               & 6 bottom encoder layers     & 16.7 & 12.2 \\ \cline{3-5} 
                         &                               & 3 bottom encoder layers     & 17.2 & 12.7 \\ \hline
\end{tabular}
\label{exp:layers}
\end{table}




\subsection{Results of utilizing extra text data}

\begin{figure}[htbp]
  \centering
  \includegraphics[width=1.0\linewidth]{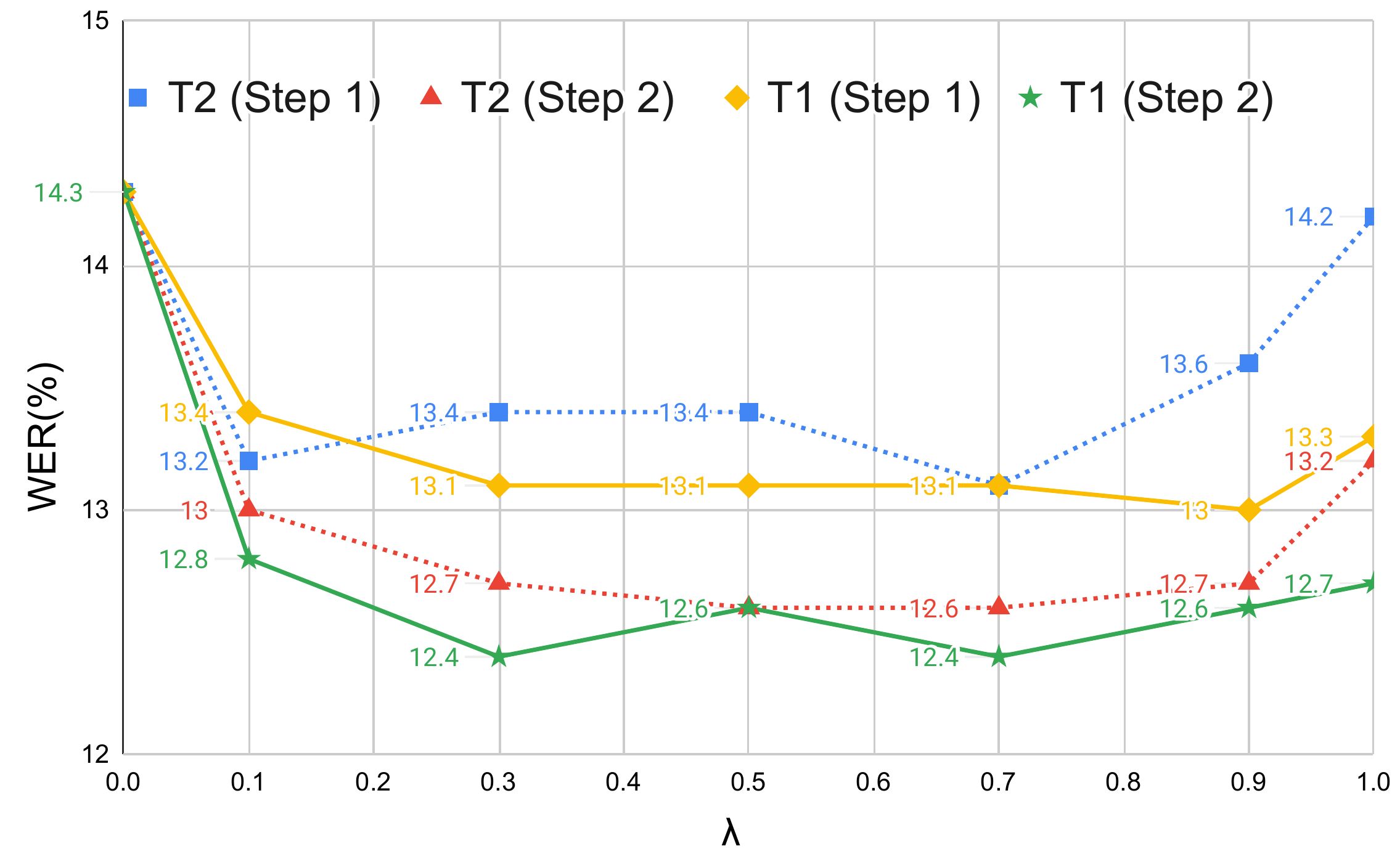}
  \caption{ The performance on the dev data after cross-lingual transfer learning when different amount of extra text data, i.e. $T1$ and $T2$, is used to fine-tune $A_3$. $\lambda$ is the interpolation factor in Equation (\ref{totalLoss}). Step 1 and Step 2 are explained in Fig. \ref{fine_tune}.}
  \label{fig:textrate}
\end{figure}

We first present the ASR performance on development data when using extra text data $T1$ and $T2$ to fine-tune $A_3$ after cross-lingual transfer learning. The results are shown in Fig. \ref{fig:textrate}. We observed that using extra text data is very effective and $T1$ produces substantial improvement over $T2$. We also observed that Step 2 (see Fig. \ref{fine_tune}), i.e. to use labeled data to fine-tune the E2E network, is essential. Finally,  $\lambda=0.7$ yields the best results in most of the cases. 

We then employ $T1$ with $\lambda=0.7$ on the test set and the result is reported in the row 6 of Table \ref{exp:results}. Using extra text data significantly improves $A_3$ by 13.6\% relative WER (from 10.3\% to 8.9\%). With the help of the extra text data, the proposed architecture outperforms the Transformer baseline by 11.9\% relative WER (from 10.1\% to 8.9\%). 
\vspace{-0.2cm}

We now investigate the effect of an external language model on the proposed architecture $A_3$. We train a Recurrent Neural Network LM (RNN-LM) as a 1-layer LSTM with 1024 cells using both transcriptions of the training data and the extra text data $T1$, then integrate the RNN-LM into inference process of $A_3$ (row 5 and 6 in Table \ref{exp:results}). Results are reported in Table \ref{exp:rescore}. As we can see, after fine-tuning with $T1$, $A_3$ still benefits from the external RNN-LM and we observed 34.8\% relative WER reduction (from 8.9\% to 5.8\%). 

\begin{table}[htbp]
\centering
\caption{ASR performance of $A_3$ (row 5 and 6 from Table \ref{exp:results}) on test set with and without RNN-LM.}
\begin{tabular}{c|c|c}
\hline
                  \begin{tabular}[c]{@{}c@{}}Row No.\\ (From Table \ref{exp:results}) \end{tabular}   & External LM & WER(\%) \\ \hline
\multirow{2}{*}{5} & -       & 10.3    \\ \cline{2-3} 
                    & + RNN-LM   & \textbf{6.8}     \\ \hline
\multirow{2}{*}{6} & -       & 8.9     \\ \cline{2-3} 
                    & + RNN-LM   & \textbf{5.8}     \\ \hline
\end{tabular}
\label{exp:rescore}
\end{table}
\vspace{-0.2cm}
\section{Conclusions}
\label{conclusions}
In this paper, we first proposed the Transformer-LSTM based architecture which not only takes advantage of the highly effective encoding capacity of the Transformer, but also benefits from the extra text data due to the LSTM-based independent language model decoder. We then examined exploiting extra text data to boost the LSTM decoder under cross-lingual transfer learning. Experimental results show that, with the help of the extra text data, the proposed architecture significantly outperforms baselines. Additionally, the proposed architecture also offers faster decoding.

\section{Acknowledgements}
This work is supported by the project of Alibaba-NTU Singa-pore Joint Research Institute. The computational work for this article was partially on resources of the National Supercomputing Centre, Singapore (https://www.nscc.sg).

\newpage
\bibliographystyle{IEEEtran}
\bibliography{mybib}

\begin{thebibliography}{10}
\providecommand{\url}[1]{#1}
\csname url@samestyle\endcsname
\providecommand{\newblock}{\relax}
\providecommand{\bibinfo}[2]{#2}
\providecommand{\BIBentrySTDinterwordspacing}{\spaceskip=0pt\relax}
\providecommand{\BIBentryALTinterwordstretchfactor}{4}
\providecommand{\BIBentryALTinterwordspacing}{\spaceskip=\fontdimen2\font plus
\BIBentryALTinterwordstretchfactor\fontdimen3\font minus
  \fontdimen4\font\relax}
\providecommand{\BIBforeignlanguage}[2]{{%
\expandafter\ifx\csname l@#1\endcsname\relax
\typeout{** WARNING: IEEEtran.bst: No hyphenation pattern has been}%
\typeout{** loaded for the language `#1'. Using the pattern for}%
\typeout{** the default language instead.}%
\else
\language=\csname l@#1\endcsname
\fi
#2}}
\providecommand{\BIBdecl}{\relax}
\BIBdecl

\bibitem{Tung2020}
V.~T. Pham, H.~Xu, K.~Yerbolat, Z.~Zeng, E.~S. Chng, C.~Ni, B.~Ma, and H.~Li,
  ``Independent language model architecture for end-to-end asr,'' in
  \emph{Proc. of ICASSP}, 2020, pp. 7054--7058.

\bibitem{Graves:2014:TES:3044805.3045089}
A.~Graves and N.~Jaitly, ``Towards end-to-end speech recognition with recurrent
  neural networks,'' in \emph{Proc. of ICML}, 2014, pp. 1764--1772.

\bibitem{Hori2018}
T.~Hori, J.~Cho, and S.~Watanabe, ``End-to-end speech recognition with
  word-based {RNN} language models,'' in \emph{Proc. of HLT}, 2018, pp.
  389--396.

\bibitem{deepAndShallowFusion}
{\c{C}}.~G{\"{u}}l{\c{c}}ehre, O.~Firat, K.~Xu, K.~Cho, L.~Barrault, H.~Lin,
  F.~Bougares, H.~Schwenk, and Y.~Bengio, ``On using monolingual corpora in
  neural machine translation,'' \emph{CoRR}, vol. abs/1503.03535, 2015.

\bibitem{ChanJLV16}
W.~Chan, N.~Jaitly, Q.~V. Le, and O.~Vinyals, ``Listen, attend and spell: {A}
  neural network for large vocabulary conversational speech recognition,'' in
  \emph{Proc. of {ICASSP}}, 2016, pp. 4960--4964.

\bibitem{Transformer2017}
A.~Vaswani, N.~Shazeer, N.~Parmar, J.~Uszkoreit, L.~Jones, A.~N. Gomez,
  L.~Kaiser, and I.~Polosukhin, ``Attention is all you need,'' in \emph{Proc.
  of NIPS}, 2017, pp. 5998--6008.

\bibitem{Huang2013}
J.~{Huang}, J.~{Li}, D.~{Yu}, L.~{Deng}, and Y.~{Gong}, ``Cross-language
  knowledge transfer using multilingual deep neural network with shared hidden
  layers,'' in \emph{Proc. of ICASSP}, 2013, pp. 7304--7308.

\bibitem{Heigold2013}
G.~{Heigold}, V.~{Vanhoucke}, A.~{Senior}, P.~{Nguyen}, M.~{Ranzato},
  M.~{Devin}, and J.~{Dean}, ``Multilingual acoustic models using distributed
  deep neural networks,'' in \emph{Proc. of ICASSP}, 2013, pp. 8619--8623.

\bibitem{Shan2019}
C.~{Shan}, C.~{Weng}, G.~{Wang}, D.~{Su}, M.~{Luo}, D.~{Yu}, and L.~{Xie},
  ``Investigating end-to-end speech recognition for mandarin-english
  code-switching,'' in \emph{Proc. of ICASSP}, 2019, pp. 6056--6060.

\bibitem{Tong2017}
S.~Tong, P.~N. Garner, and H.~Bourlard, ``Multilingual training and
  cross-lingual adaptation on ctc-based acoustic model,'' \emph{CoRR}, vol.
  abs/1711.10025, 2017.

\bibitem{Dalmia2018}
S.~Dalmia, R.~Sanabria, F.~Metze, and A.~W. Black, ``Sequence-based
  multi-lingual low resource speech recognition,'' \emph{CoRR}, vol.
  abs/1802.07420, 2018.

\bibitem{speechTrans}
L.~Dong, S.~Xu, and B.~Xu, ``Speech-transformer: A no-recurrence
  sequence-to-sequence model for speech recognition,'' in \emph{Proc. of
  ICASSP}, 2018, pp. 5884--5888.

\bibitem{TransASRU}
S.~Karita, X.~Wang, S.~Watanabe, T.~Yoshimura, W.~Zhang, N.~Chen, T.~Hayashi,
  T.~Hori, H.~Inaguma, Z.~Jiang, M.~Someki, N.~Yalta, and R.~Yamamoto, ``A
  comparative study on {Transformer} vs {RNN} in speech applications,'' in
  \emph{Proc. of ASRU}, 2019, pp. 449--456.

\bibitem{BaKH16}
L.~J. Ba, J.~R. Kiros, and G.~E. Hinton, ``Layer normalization,'' \emph{CoRR},
  vol. abs/1607.06450, 2016.

\bibitem{HeZRS16}
K.~He, X.~Zhang, S.~Ren, and J.~Sun, ``Deep residual learning for image
  recognition,'' in \emph{Proc. of CVPR}, 2016, pp. 770--778.

\bibitem{combineTransformerLSTM}
M.~X. Chen, O.~Firat, A.~Bapna, M.~Johnson, W.~Macherey, G.~Foster, L.~Jones,
  N.~Parmar, M.~Schuster, Z.~Chen, Y.~Wu, and M.~Hughes, ``The best of both
  worlds: Combining recent advances in neural machine translation,''
  \emph{CoRR}, vol. abs/1804.09849, 2018.

\bibitem{TangMRS18}
G.~Tang, M.~Müller, A.~Rios, and R.~Sennrich, ``Why self-attention? a targeted
  evaluation of neural machine translation architectures.'' in \emph{Proc. of
  EMNLP}, 2018, pp. 4263--4272.

\bibitem{zhang-Etal:2018:ACL2018accelerating}
B.~Zhang, D.~Xiong, and J.~Su, ``Accelerating neural transformer via an average
  attention network,'' in \emph{Proc. of ACL}, 2018, pp. 1789--1798.

\bibitem{DBLP:conf/interspeech/KoPPK15}
T.~Ko, V.~Peddinti, D.~Povey, and S.~Khudanpur, ``Audio augmentation for speech
  recognition,'' in \emph{Proc. of {INTERSPEECH}}, 2015, pp. 3586--3589.

\bibitem{koh2019building}
J.~X. Koh \emph{et~al.}, ``Building the {Singapore} {English} national speech
  corpus,'' in \emph{Proc. of INTERSPEECH}, 2019, pp. 321--325.

\bibitem{watanabe2018espnet}
S.~Watanabe, T.~Hori, S.~Karita, T.~Hayashi, J.~Nishitoba, Y.~Unno, N.~E.~Y.
  Soplin, J.~Heymann, M.~Wiesner, N.~Chen \emph{et~al.}, ``Espnet: End-to-end
  speech processing toolkit,'' in \emph{Proc. of INTERSPEECH}, 2018, pp.
  2207--2211.

\bibitem{kim2017joint}
S.~Kim, T.~Hori, and S.~Watanabe, ``Joint {CTC}-attention based end-to-end
  speech recognition using multi-task learning,'' in \emph{Proc. of ICASSP},
  2017, pp. 4835--4839.

\end{thebibliography}
\end{document}